\documentclass[9pt]{article}
\usepackage{spconf,amsmath,graphicx}
\usepackage{booktabs}       
\usepackage{xspace}
\usepackage{afterpage}
\usepackage{float}
\usepackage{wrapfig} 

\def\x{{\mathbf x}}
\def\z{{\mathbf z}}

\title{Voice Impersonation using Generative Adversarial Networks}
\name{Yang Gao\sthanks{Thanks to Raymond (Yangyang) Xia for his suggestions and advice}, Rita Singh\sthanks{This work was funded by Schmidt Sciences, Palo Alto, CA.}, Bhiksha Raj}
\address{Electrical and Computer Engineering Department, Carnegie Mellon University \\ yanggao,rsingh,bhiksha@cs.cmu.edu}
\begin{document}
\ninept
\maketitle
\begin{abstract}

Voice impersonation is not the same as voice transformation, although
the latter is an essential element of it. In voice impersonation,
the resultant voice must convincingly convey the impression of having
been naturally produced by the target speaker, mimicking not only the
pitch and other perceivable signal qualities, but also the \textit{style} of
the target speaker. In this paper, we propose a novel neural-network
based speech quality- and style-mimicry framework for the synthesis of
impersonated voices.  The framework is built upon a fast and accurate
generative adversarial network model. Given spectrographic
representations of source and target speakers' voices, the model
learns to mimic the target speaker's voice quality and style,
regardless of the linguistic content of either's voice, generating a synthetic spectrogram from which the 
time-domain signal is reconstructed using the Griffin-Lim method. In effect, this model
reframes the well-known problem of style-transfer for images as the
problem of style-transfer for speech signals, while intrinsically
addressing the problem of durational variability of speech
sounds. Experiments demonstrate that the model can generate extremely
convincing samples of impersonated speech. It is even able to
impersonate voices across different genders effectively. Results are
qualitatively evaluated using standard procedures for evaluating
synthesized voices.


\end{abstract}
\begin{keywords}
Voice impersonation, generative adversarial network, style transformation, style transfer
\end{keywords}

\section{Introduction}
\label{sec:intro}
Voice impersonation, where one person attempts to mimic the voice of another to sound like the other person, is a complex phenomenon. 
It is often performed by the impersonator by intuitively trying to match the most prominent aspects of the voice of the person mimicked. The most common sets of characteristics in voice that are varied by the impersonator in this process are elements of \textit{voice quality} and elements of {\em style}. Voice quality, however, is a hard-to-describe entity. While some aspects of it, such as nasality, roughness, breathiness etc. have been identified in the literature, most of these are not quantifiable in a manner that permits meaningful comparison across speakers. Elements of \textit{style}, which include temporal pitch and energy patterns that are characteristic of the target speaker's cadence, speaking rate, prosody etc. are similarly ambiguous in terms of their exact definition, quantifiability and comparability. In addition to quality and style, idiosyncrasies of language such as the variants of grammar, choice of vocabulary, usage of words such as ``um'', ``uh'' etc. are also mimicked by the impersonator to create an impression of the target speaker's voice; these are however currently not the focus of this paper.

In this paper, we attempt to address the problem of automatic generation of impersonations by transferring the stylistic content of speech from the target speaker to that of the impersonator. The objective is to convert only specific aspects of the voice, without modifying other content. Ideally, we would be able to identify, isolate and explicitly measure the aspects of voice that characterize stylistic features, and then use them to modify the target voice. However, this is generally difficult because such features are subjectively or even objectively identifiable, but not quantifiable. To misquote the Hon. Potter Stewart, we may not know how to describe them, but we know when they occur.  

Our objective is to develop a mechanism to transfer such identifiable (albeit unquantifiable) aspects of style. 
In the absence of mechanisms to isolate, quantify and measure these, generative approaches such as source-filter models within which explicit components of the source of filter can be modified \cite{erro2013parametric}, spectral-transformation models such as PSOLA \cite{valbret1992voice}, etc. become inappropriate. Instead, we must use an alternate approach, where the ability to transform the voice must be learned solely from the identifiability of the target characteristics. 

Prior research that is of greatest relevance in this context relates to {\em voice transformation}, which deals with the specific problem of converting a source voice into a target one. Voice transformation has had a long history, and at the surface addresses some of the issues we mention.
Conventionally, voice transformation modifies the instantaneous characteristics of a source signal, such as pitch \cite{rao2010} and spectral envelope. The strategies used range from simple codebook-based conversion \cite{abe90} and minimum-mean-squared error linear estimators \cite{AlanB} to sophisticated neural network models \cite{chen2014voice}. While these methods are all frequently quite effective at transforming instantaneous characteristics of the signal, 
and can even map some prosodic cues, they are generally insufficient to capture unmeasurable, unquantifiable {\em style} in the more general sense of the word. When trained, they are heavily reliant on the availability of {\em parallel} recordings of the source and target speaker saying the same utterances, providing exact examples of what is considered ideal conversion. In most cases, in order to learn the voice conversion effectively, these recordings must also be perfectly time aligned, a requirement that is generally satisfied by time-warping the recordings to align them to one another. Realization of the {\em hard} targets required to learn the conversion is not only unrealistic, the alignment required may also be fundamentally inappropriate when the objective is not to learn to perform wholesale conversion of voice, but only to transform style.

In this context, recent advances in the science of learning generative models provide us new directions. Rather than attempting to learn a mapping between parallel signals, 
the new models attempt to {\em discriminate} instead between data that do have the desired (identifiable but possibly unquantifiable) stylistic feature(s), and those that do not. Generators that attempt to produce data with any specific characteristic(s) must now learn to do so such that they ``fool'' the discriminator. Since the features are unquantifiable, the discriminator itself must, in fact also be learned. Both the generators and the discriminators are modeled by deep neural networks, which are known to be able to model any transformation with appropriate design and sufficient training data. Since the primary driver of the learning process is discrimination, parallel data such as those needed for conventional voice-conversion methods are not required.

These \textit{Generative Adversarial Networks}, or GANs have been very successfully applied to a variety of problems in image generation \cite{GAN_original}, learning feature representations \cite{radford2015unsupervised} and  \textit{style transfer} \cite{DiscoGAN,CycleGAN,Can,ImageTranslation,gatys2016image}, wherein the algorithms involved result in fast and vivid generation of images of different artistic styles ranging from simple photographs to painting styles of selected artists. In our work, we harness the power of these models for the problem of style transfer in speech. 

At the outset, we note that speech signals have several problems that are not inherent to images. Unlike images, speech sounds are not of fixed size (i.e., not fixed in duration), and lose much of their stylistic characteristics when they are scaled down to be so. Generation of time-series data such as speech is also a more challenging problem compared to images. Naive implementations of the process may result in generation of data that have lost linguistic, stylistic or even intelligible content. In this work, we propose multiple GAN models for the problem of voice transformation. Our models, and their corresponding learning algorithms, are designed to consider the specific challenges inherent in speech. Specifically, we show how, by appropriate choice of model structure and learning algorithm, and by introducing the appropriate discriminators in the GAN framework, {\em specific} characteristics of the voice might be retained without modifying others or losing linguistic content, in order to emulate different aspects of impersonation or voice mimicry. 

In Section \ref{sec:gan} we briefly outline the concept of GANs. In Section \ref{sec:voicegan} we describe our designs of GANs for voice modifications. In Section \ref{sec:expts} we present experimental evaluations of the proposed models and conclude with discussions in Section \ref{sec:discuss}.

\section{Generative Adversarial Networks}
\label{sec:gan}

In spite of their rather short history, GANs \cite{GAN_original} are already quite well known.  We briefly summarize their key features here, in order to set the background for the rest of the paper.

\subsection{The basic GAN model}
The Generative Adversarial Network is a generative model which, at its foundation, is a generative model for a data variable. The model is intended to generate samples that closely match draws from the actual distribution of the data.  These models differ from conventional generative models in a fundamental way in the manner in which they are learned. Conventional generative models are trained through likelihood maximization criteria, such that some (empirical estimate of the) divergence measure between the synthetic distribution encoded by the generative models, and the true distribution of the data, is minimized. In contrast, GANs are trained discriminatively, such that samples generated from the model cannot be distinguished from actual draws from the true distribution of the data.

Consider any random variable $\x$ with a probability distribution $P_\x$ that is unknown, but from which samples may be drawn. For instance, $\x$ may represent images of a particular class, samples of which may be readily available, but their actual distribution may be unknown. The GAN attempts to generate samples of $\x$ that are indistinguishable from actual samples drawn from the true distribution.
The original GAN model \cite{GAN_original} comprises a generator $G(\z)$ and discriminator $D(\x)$. The generator $G$ takes as input a random variable $\z$ drawn from some standard probability distribution function $P_\z$, {\em e.g.} a standard Normal distribution, and produces an output vector $\x_\z$. 

The discriminator $D()$ attempts to discriminate between samples $\x \sim P_\x$ that are drawn from $P_\x$, the true (but unknown) distribution we aim to model, and samples produced by the Generator $G$.  Let $T$ represent the event that a vector $\x$ was drawn from $P_\x$. The discriminator attempts to compute the {\em a posteriori} probability of $T$, {\em i.e.} $D(\x) = P(T|\x)$.

\begin{figure}[htb]
\begin{minipage}[b]{0.5\linewidth}
  \centering
  \centerline{\includegraphics[height=1.9cm]{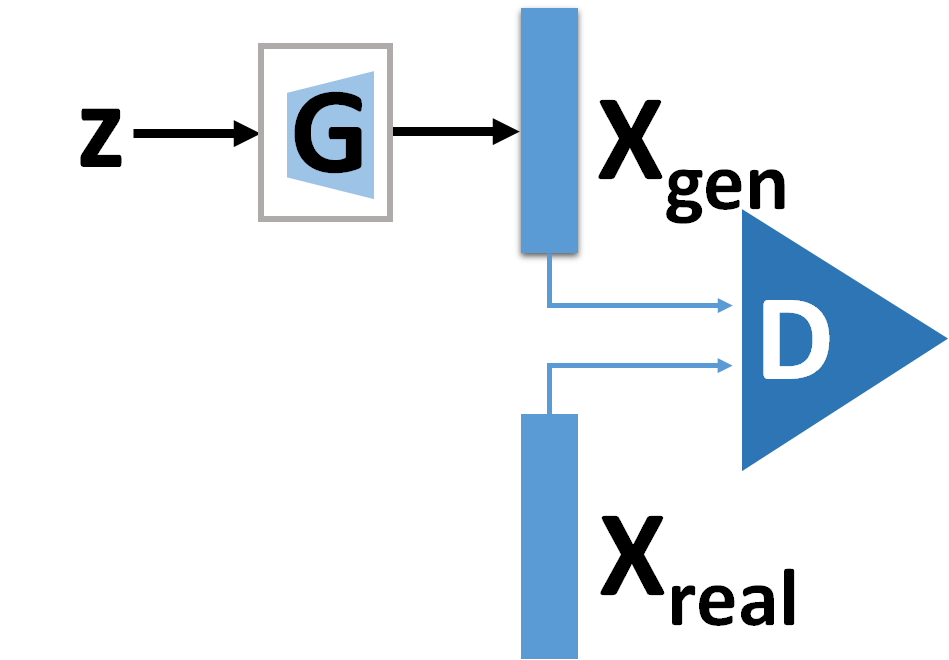}}
\caption{The original GAN model}
\label{fig:basicgan}
\end{minipage}
\begin{minipage}[b]{0.5\linewidth}
  \centering
  \centerline{\includegraphics[height=2cm]{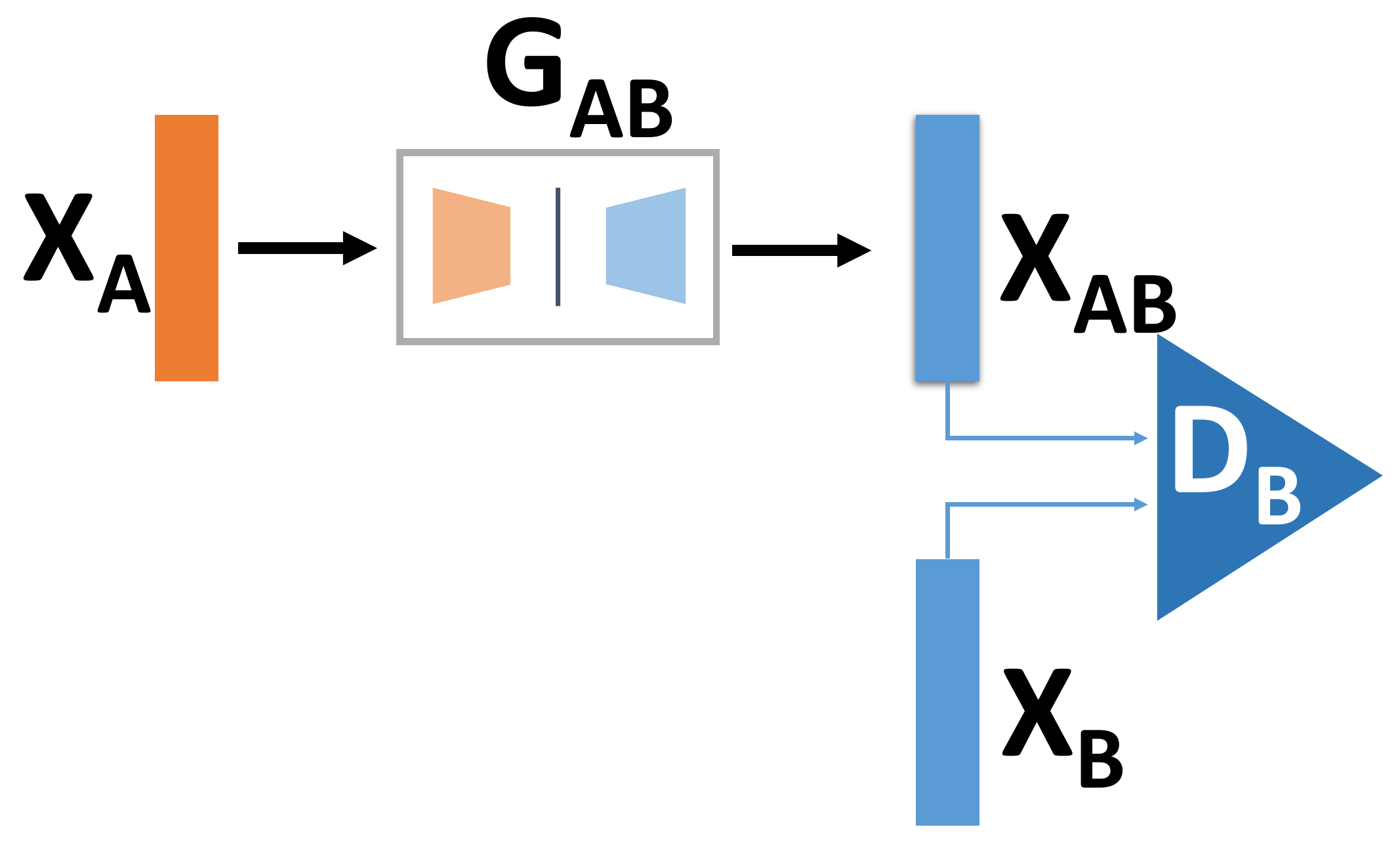}}
\caption{Style transfer by GAN}
\label{fig:res2}
\end{minipage}

\end{figure}

To train the GAN, we attempt to learn $G$ such that $D(\x_\z)$, the score output by the discriminator in response to productions by $G$ is maximized (i.e. $G$ ``fools'' the discriminator). At the same time we attempt to learn $D$ such that $D(\x_\z)$ is minimized, while also maximizing $D(\x)$ for any $\x \sim P_\x$. All of these objectives can be concurrently achieved through the following optimization:
\[
\min_G \max_D E_{\x\sim P_\x}[\log D(\x)]  + E_{z\sim P_\z}[\log(1-D(x_\z))]  
\]
The GAN training framework is illustrated in Figure \ref{fig:basicgan}.

\subsection{GANs for style transfer}
The basic GAN has been extended in a number of ways in the literature \cite{DiscoGAN,CycleGAN,Can,ImageTranslation,gatys2016image}, particularly in the context of style transfer among images, \textit{e.g.} as in Figure \ref{fig:res2}. The common underlying denominator in all of these models is that an input data instance (usually an image) $x_A$ drawn from a distribution $P_A$ is {\em transformed} to an instance $x_{AB}$ by a generator (more aptly called a ``transformer''), $G_{AB}$. The aim of the transformer is to convert $x_A$ into the style of the variable $x_B$ which natively occurs with the distribution $P_B$. 

The discriminator $D_B$ attempts to distinguish between genuine draws of $x_B$ from $P_B$ and instances $x_{AB}$ obtained by transforming draws of $x_A$ from $P_A$.
The actual optimization is achieved as follows. We define
\begin{align} 
L_G &=  E_{x_A\sim P_A}[\log (1 - D_B(x_{AB}))]  \nonumber \\
L_D &= -E_{x_B\sim P_B}[\log D_B(x_B)] - E_{x_A\sim P_A}[\log(1-D_B(x_{AB})]   \label{eq2}
\end{align}
To train the GAN, its two components are alternately updated by minimizing the two losses in Equation \ref{eq2}. The generator $G$ is updated by minimizing the ``generator loss'' $L_G$, while the discriminator is updated to minimize the ``discriminator loss'' $L_D$.

\begin{figure}[htb]
  \centering
  \centerline{\includegraphics[width=0.8\columnwidth]{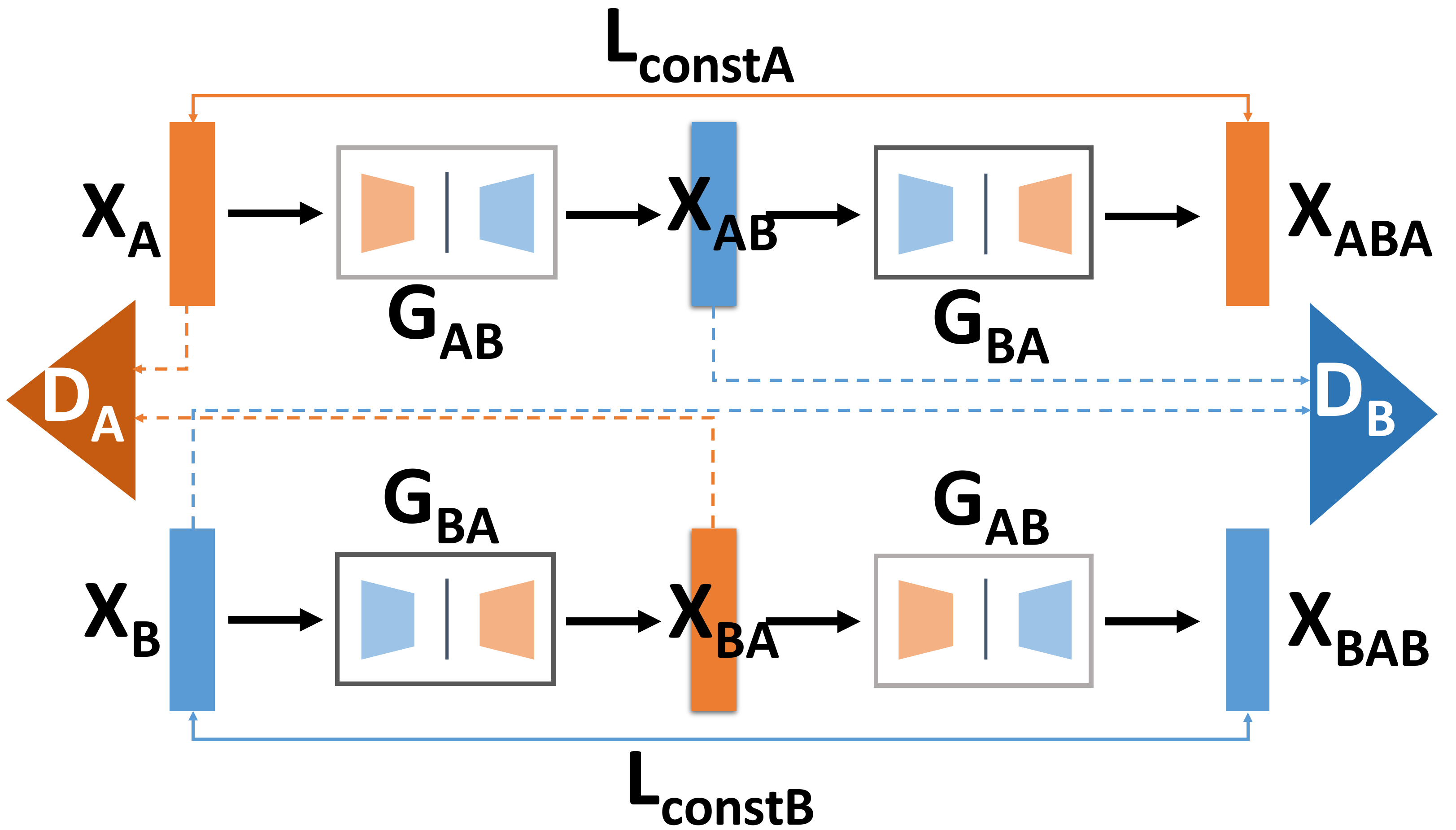}}
\caption{The DiscoGAN model}
\label{fig:discores}
\end{figure}

Our work is however more directly based on the ``DiscoGAN'' model \cite{DiscoGAN}, shown in Figure \ref{fig:discores}. The DiscoGAN is a symmetric model which attempts to transform two categories of data, $A$ and $B$, into each other. The DiscoGAN includes two generators (more aptly called ``transformers'') $G_{AB}$ and $G_{BA}$.  $G_{AB}$ attempts to transform any draw $x_A$ from the distribution $P_A$ of $A$ into $x_{AB} = G_{AB}(x_A)$, such that $x_{AB}$ is indistinguishable from draws $x_B$ from the distribution $P_B$ of $B$. $G_{BA}$ does the reverse.

It also includes two discriminators, $D_A$ and $D_B$.  $D_A$ attempts to discriminate between actual draws from $P_A$ and draws from $P_B$ that have been transformed by $G_{BA}$, and $D_B$ performs the analogous operations for draws from $P_B$. The generators and discriminators must all be jointly trained.

The training process for the DiscoGAN is similar to that for the model in Figure \ref{fig:res2}, with one significant modification: in addition to the losses that emphasize the competition between the generators and the discriminators, we now include the requirement that $G_{AB}$ and $G_{BA}$ must be inverses of each other to the extent possible, {\em i.e.} for any $x_A$ from $A$, $x_{ABA} = G_{BA}(G_{AB}(x_A))$ must be close to the original $x_A$, and similarly for any $x_B$ from $B$, $x_{BAB} = G_{AB}(G_{BA}(x_B))$ must be close to the original $x_B$. This requirement is encoded through two reconstruction losses $L_{CONST_A}$ and $L_{CONST_B}$ where
\begin{equation} \label{eq3}
L_{CONST_A} = d(G_{BA}(G_{AB}(x_A)), x_A)    
\end{equation}
and $L_{CONST_B}$ is symmetrically defined. The generator loss for $G_{AB}$ is defined as:
\begin{equation} \label{eq4}
L_{GAN_{AB}} =  L_{CONST_A} + L_{G_B}      
\end{equation}
where $L_{G_B}$ is defined as in Equation \ref{eq2}.  We define the generator loss for $G_{AB}$ in a symmetric manner. The overall generator loss is $L_G = L_{GAN_{AB}} + L_{GAN_{BA}}$. The discriminator loss $L_D$ is defined as $L_D = L_{D_A} + L_{D_B}$, $L_{D_A}$ and $L_{D_B}$ are defined as in Equation \ref{eq2}. Finally, in the implementation of DiscoGAN \cite{DiscoGAN}, a \textit{feature loss} is also added to compare the feature similarity between the generated data and the real data.  As before, the generators and discriminators are trained by alternate minimization of the generator and discriminator losses.
%
%
%
%
%
\section{GANs for voice mimicry}
\label{sec:voicegan}

The DiscoGAN was originally designed to transform style in images. In order to apply the model to speech, we first convert it to an invertible, picture-like representation, namely a spectrogram. We operate primarily on the {\em magnitude} spectrogram, retaining the phase of input signals to be transformed, to recreate the transformed signals from the transformed magnitude spectrogram.

But first, we must make several key modifications to the DiscoGAN model. The original DiscoGAN was designed to operate on images of fixed size. For it to work with inherently variable-sized speech signals, this constraint must be relaxed in its new design.  Secondly, it is important to ensure that the {\em linguistic} information in the speech signal is not lost, even through the signal itself is modified. Sufficient constraints must be added to the model for this. Finally, since our objective is to modify specific aspects of the speech, \textit{e.g.} style, we must add extra components to our model to achieve this. We call our model, which incorporates all these modifications, the {\em VoiceGAN}.

\subsubsection{Retaining Linguistic Information}
\label{sssec:subsubhead}
Linguistic information is encoded largely in the details of the spectral envelope. To ensure that this is retained, we modify our reconstruction loss as: 
\begin{equation} \label{eq5}
L_{CONST_A}  = \alpha d(x_{ABA}, x_A) + \beta d(x_{AB}, x_A)    
\end{equation}
Here, the term $d(x_{AB}, x_A)$ attempts to retain the \textit{structure} of $x_A$ even after it has been converted to $x_{AB}$. Careful choice of $\alpha$ and $\beta$ ensures both, accurate reconversion and retention of linguistic information, after conversion to $x_{AB}$.

\subsubsection{Variable-length Input Generator and Discriminator}
\label{sssec:subsubhead}
\begin{figure}[htb]
\begin{minipage}[b]{.45\linewidth}
  \centering
  \centerline{\includegraphics[width=3.5cm]{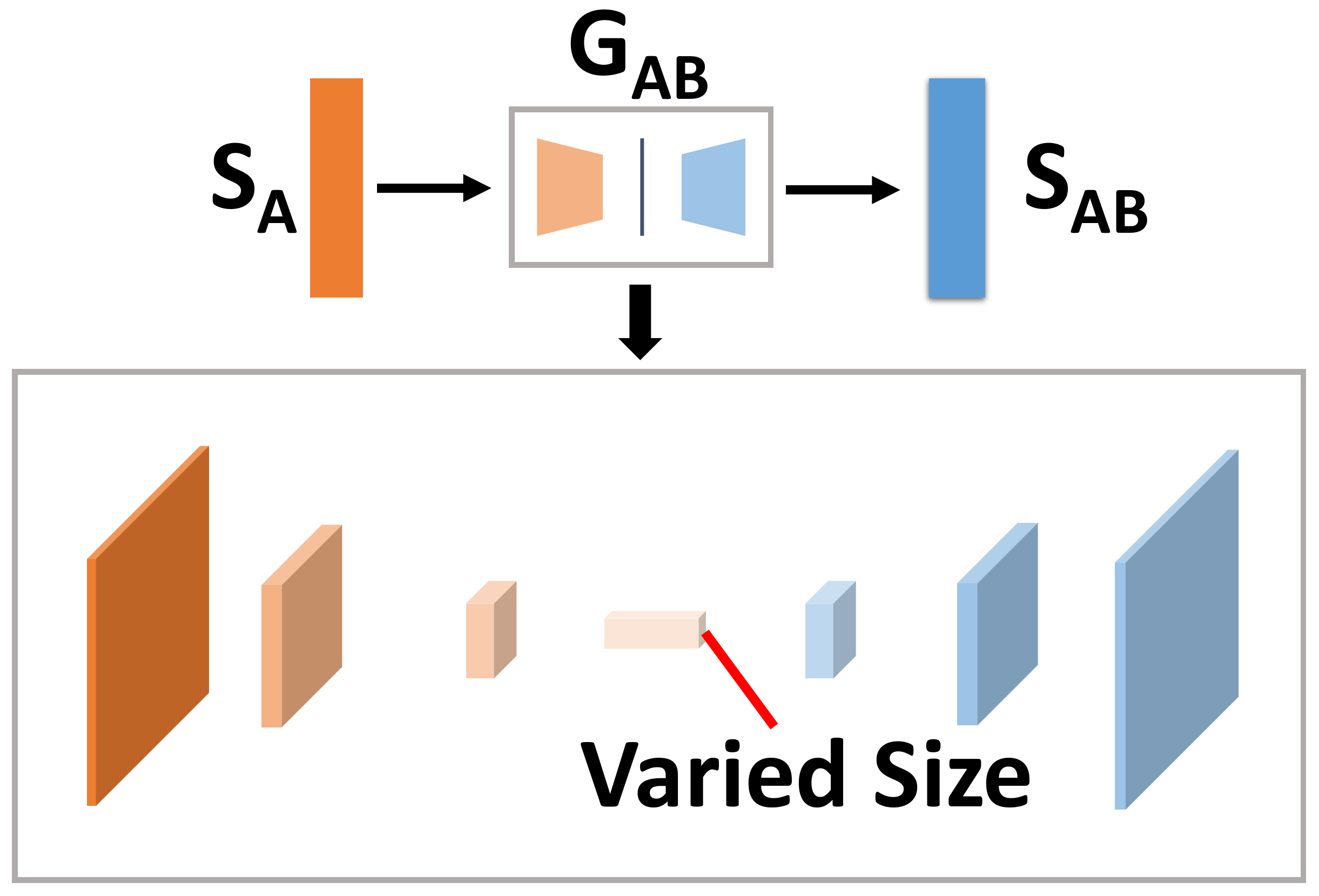}}
  \centerline{(a) Generator $G_A$}\medskip
\end{minipage}
\hfill
\begin{minipage}[b]{0.53\linewidth}
  \centering
  \centerline{\includegraphics[width=\columnwidth]{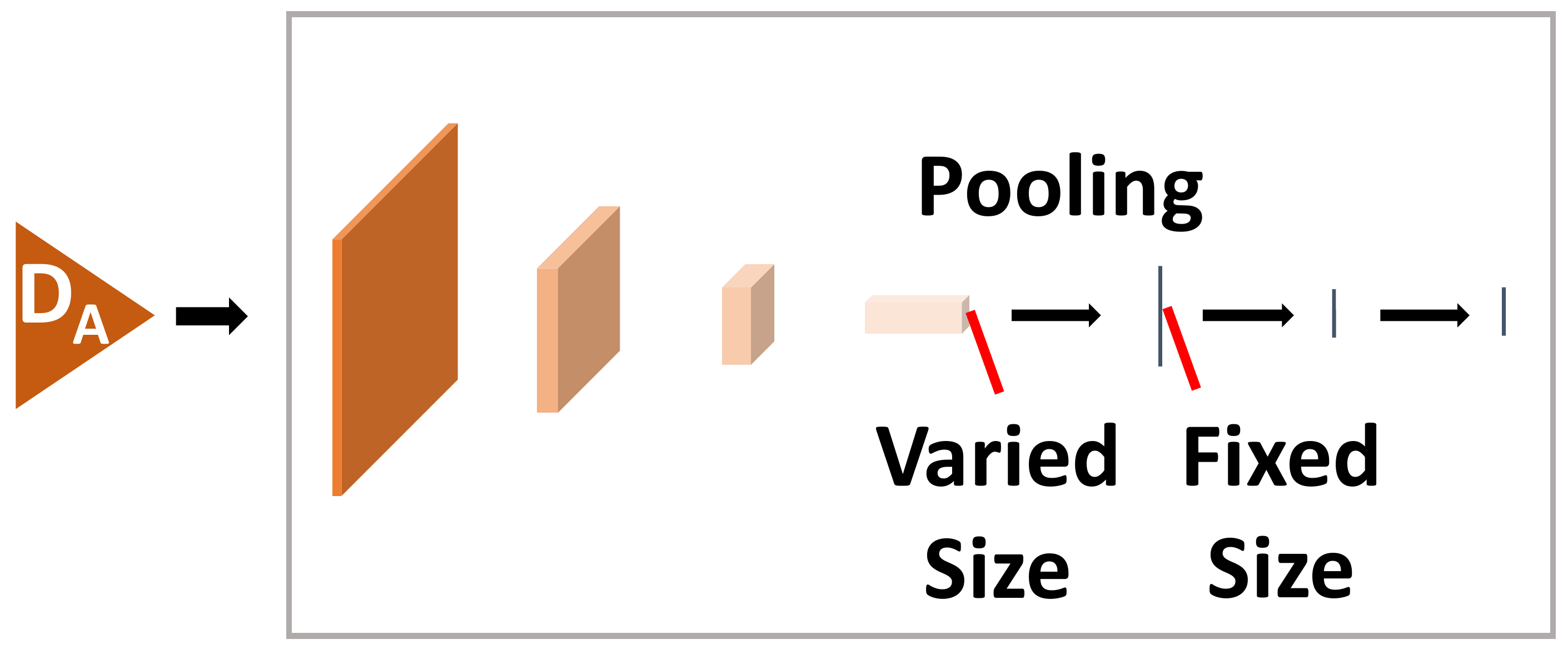}}
  \centerline{(b) Discriminator $D_A$}\medskip
\end{minipage}
\caption{Visualization of Generator $G_A$ and Discriminator $D_A$. Architectures of other counterparts are similar in structure. The number of convolutional layers is larger in the actual implementation.}
\label{fig:varsize}
\end{figure}
To account for the fact that unlike images, speech signals are of variable length that cannot be scaled up or down, we must make modifications to the generators and discriminators. The modified structures are shown in Figure \ref{fig:varsize}. Figure \ref{fig:varsize} (a) shows the structure of the original generator in DiscoGAN. Based on its fully convolutional structure, it can handle variable length inputs. Figure \ref{fig:varsize} (b), we shows the architectural details for our proposed discriminator in VoiceGAN. In this, an adaptive pooling layer is added after the CNN layers, and before the fully connected layer. It includes channel-wise pooling in which each channel's feature map is pooled into a single element. This converts any variable-sized feature map into a vector of a fixed number of dimensions, with as many components as the number of channels. 

\subsubsection{Style Embedding Model ($D_S$)}
\label{sssec:stylemodel}
\begin{figure}[htb]
\begin{minipage}[b]{1.0\linewidth}
  \centering
  \centerline{\includegraphics[width=0.78\columnwidth]{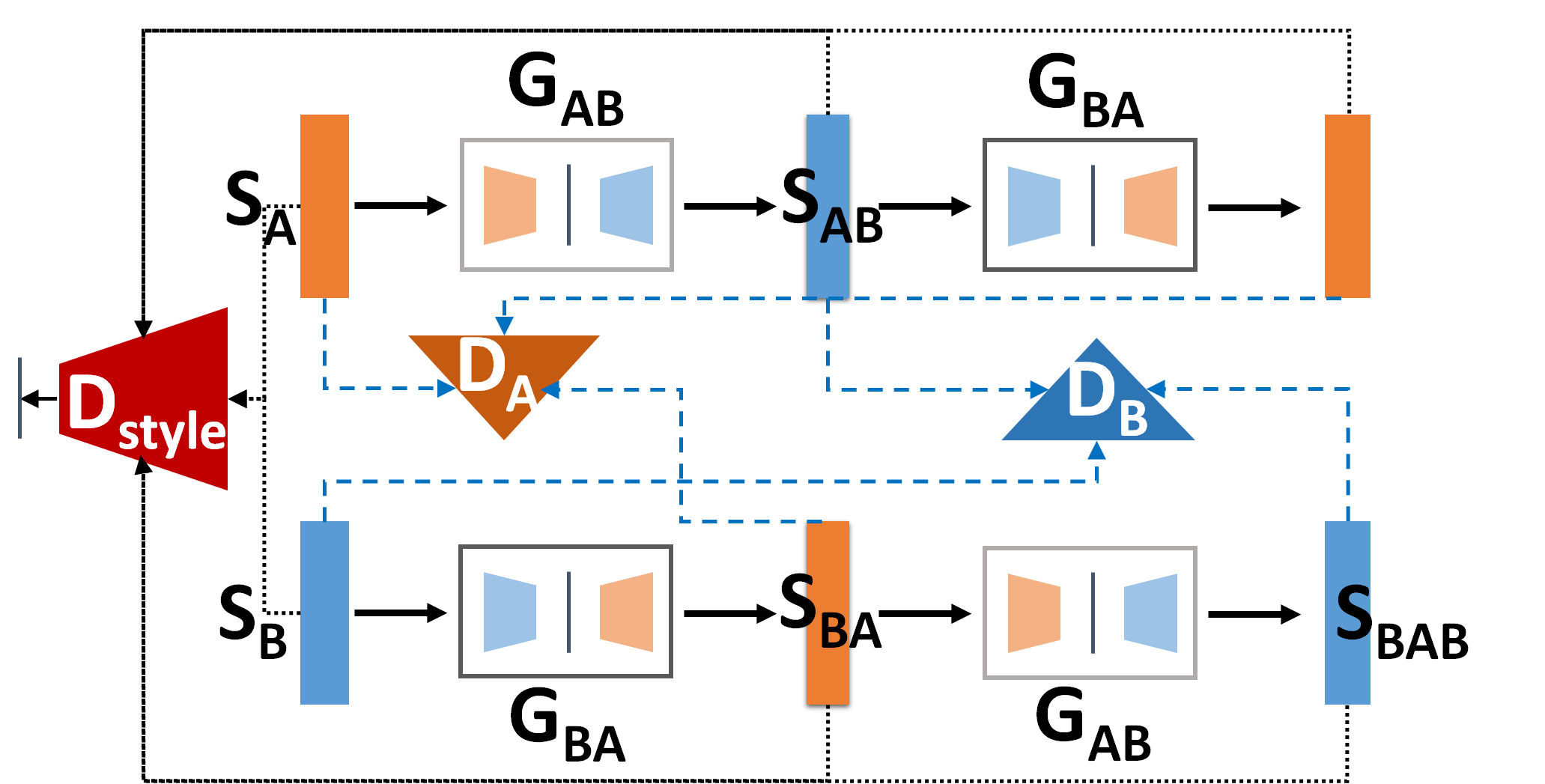}}
\end{minipage}
\caption{The proposed VoiceGAN Model}
\label{fig:voicegan1}
\end{figure}

In addition to the discriminator that distinguishes between the generated data and real data, we add a second type of discriminator  to our model to further extract the target \textit{style} information from input data and to make sure that the generated data still has this style information embedded in it. To achieve this, we include a discriminator $D_S$ that is similar in architecture to that in Figure \ref{fig:voicegan1}.



The discriminator $D_S$ determines if the original and transformed signals match the desired style. To do so we introduce the following style loss:
\begin{equation} \label{eq6}
\begin{split}
L_{D_{STYLE-A}}=& d(D_S(x_A), label_A) + d(D_S(x_{AB}), label_B) \\
        &+ d(D_S(x_{ABA}), label_A)         
\end{split}
\end{equation}
\begin{equation} \label{eq7}
L_{D_{STYLE}} = L_{D_{STYLE-A}} + L_{D_{STYLE-B}}  
\end{equation}
Note that the style loss could include multiple discriminators for multiple aspects of style. 

\subsubsection{Total Loss}
\label{sssec:totalloss}
Our final training objectives to be minimized for the generator and discriminator are represented by $L_G$ and $L_D$ respectively as follows:

\begin{equation}\label{eq8}
\begin{split}
L_G &= L_{GAN_{AB}} + L_{GAN_{BA}} \\
    &= L_{G_B} + L_{CONST_A} + L_{G_A} + L_{CONST_B}
\end{split}
\end{equation}

\begin{equation} \label{eq9}
L_D = L_{D_A} + L_{D_B} + L_{D_{STYLE}}      
\end{equation}

\section{Experiments and Results}
\label{sec:expts}
We use the TIDIGITS \cite{tid} dataset. This dataset comprises a total of 326 speakers: 111 men, 114 women, 50 boys and 51 girls. Each speaker reads 77 digit sentences. The sampling rate of the audio is 16000 Hz. We chose to use this database due to its relatively simple linguistic content. For the purpose of demonstration, we choose an unquantifiable, but identifiable characteristic: gender. Our goal then is to show that these data can be used to \textit{learn} to convert the gender of a speaker's voice. In the discussion below, therefore, ``style'' refers to gender. We note that any other characteristic may have been similarly chosen.

\subsection{Model implementation}
\label{ssec:modelimp}


The model architecture is that of the VoiceGAN described above. The generator network in the model comprises a 6-layer CNN encoder and a 6-layer transposed CNN decoder. The discriminator network comprises a 7-layer CNN with adaptive pooling. We employ batch normalization \cite{BN} and leaky ReLU activations \cite{LeakyReLU} in both the networks. The number of filters in each layer is an increasing power of 2 (32, 64, 128). When training the networks, a smoothness constraint, comprising the cumulative first order difference between adjacent columns in the spectrogram, is added to the loss to enhance the temporal continuity of the generated spectrogram. Results are available at \cite{examples}.

%

\begin{figure}[htb]

\begin{minipage}[b]{1.0\linewidth}
  \centering
  \centerline{\includegraphics[width=8.5cm]{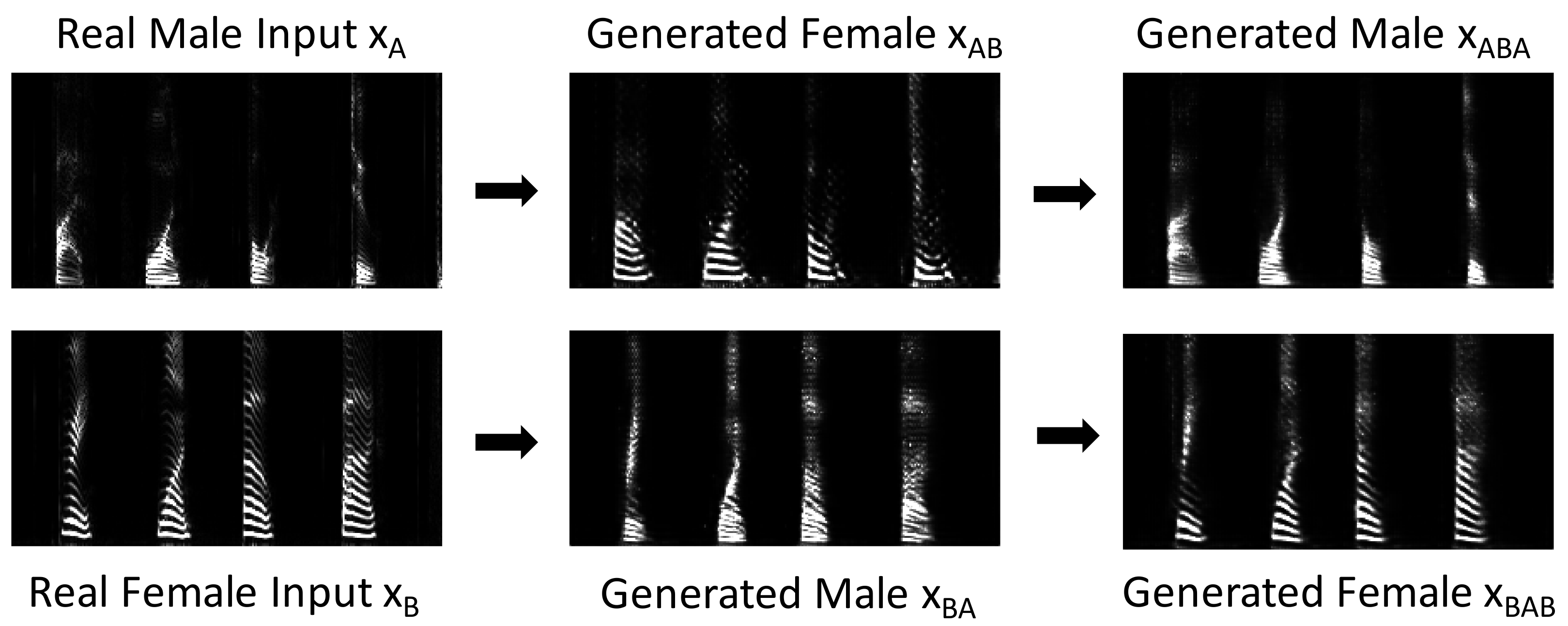}}
\end{minipage}

\caption{Visualization of spectrograms generated from a speaker saying ``3 1 oh 5'' (first row) and ``5 1 4 2'' (second row). For each spectrogram, frequencies on the y-axis range from 0-4 kHz.}
\label{fig:res}
\end{figure}


\subsection{Quality evaluation of generated results}
\label{ssec:subhead}

\subsubsection{Style Classification Test}
\label{sssec:subsubhead}
We use an independently-trained CNN-based classifier to predict the style of our generated data. 
The classifier was trained on 800 utterances from speakers of both genders. The results show that 100\% of the generated data are classified as the target speaker's style, which indicates that our VoiceGAN network achieves good style transfer performance. 

\subsubsection{Speech Signal to Noise Ratio (SNR) Test}
\label{sssec:subsubhead}

To evaluate the quality of our generated speech signal \cite{liu2006assessment}, we also conduct a signal-to-noise (SNR) ratio test using the standard NIST STNR method and the WADA SNR method \cite{WADA}. The results are shown in Table 2. For each data class, we randomly select 40 samples from our test dataset (20 for each speaker) and compute the mean and variance of the generated results. The WADA test results are all around 100 dB since our generated noise is not well-modeled by Gaussian noise. The STNR test results show that our generated data is of good quality. For evaluation, the time-domain signal is reconstructed from the generated spectrogram using the Griffin-Lim method, which is based on an iterative procedure that minimizes the mean square error between the modified magnitude spectrogram and the actual signal’s spectrogram. Details of this method are explained in \cite{GL2}. We find that the Griffin-Lim method does not reduce the voice quality to any significant degree.

\begin{table}[t]
  \caption{NIST STNR TEST} 
  
  \label{tab:symb_ls}
  \centering
  \begin{tabular}{lll}
    \toprule
    Data (use GL-method) & A ($dB$) & B ($dB$) \\
    \midrule
    Original signal & 55.60$\pm$4.97 & 52.91$\pm$3.58\\
    $X_A$ and $X_B$ & 54.97$\pm$6.28 & 52.15$\pm$3.70\\
    $X_{AB}$ and $X_{BA}$ & 49.64$\pm$1.80 & 49.92$\pm$4.36\\
    $X_{ABA}$ and $X_{BAB}$ & 53.58$\pm$2.69 & 50.05$\pm$2.12\\
    \bottomrule
  \end{tabular}
\end{table}

\vspace{-0.1in}
\section{Conclusions}
\label{sec:discuss}

The VoiceGAN model is observably able to transfer style from one speaker to another. As proposed however, this model remains vanilla and many extensions are possible. The method is easily extended to other stylistic features that may be identified. In principle, while longer-term prosodic-level style features may also be transferred, simple binary discriminators may no longer be useful for such characteristics. More continuous-valued discrimination may be required. We have not verified if multiple style aspects may be \textit{concurrently} modified. These remain areas of ongoing research. In preliminary experiments we have verified that even {\em linguistic} content may be modified if we so choose; however doing so in a measurable and controlled manner is a challenge that remains to be addressed.
Future versions of the VoiceGAN model will continue to incorporate the most relevant innovations in the area of adversarial modeling.

\clearpage
\newpage
\bibliographystyle{IEEEbib}
\bibliography{strings,refs}

\end{document}